**PROCEEDINGS OF THE ROYAL SOCIETY A**

rspa.royalsocietypublishing.org# Collective behaviour of large number of vortices in the plane

Yuxin Chen, Theodore Kolokolnikov and Daniel Zhirov

Department of Mathematics and Statistics, Dalhousie University, Halifax, Canada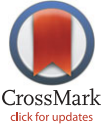

**Research**

**Cite this article:** Chen Y, Kolokolnikov T, Zhirov D. 2013 Collective behaviour of large number of vortices in the plane. Proc R Soc A 469: 20130085.
http://dx.doi.org/10.1098/rspa.2013.0085

Received: 8 February 2013
Accepted: 22 May 2013

**Subject Areas:**
differential equations, applied mathematics, biophysics

**Author for correspondence:**
Theodore Kolokolnikov
e-mail: tkolokol@mathstat.dal.caWe investigate the dynamics of $N$ point vortices in the plane, in the limit of large $N$. We consider *relative equilibria*, which are rigidly rotating lattice-like configurations of vortices. These configurations were observed in several recent experiments. We show that these solutions and their stability are fully characterized via a related *aggregation model* which was recently investigated in the context of biological swarms. By using this connection, we give explicit analytical formulae for many of the configurations that have been observed experimentally. These include configurations of vortices of equal strength; the $N+1$ configurations of $N$ vortices of equal strength and one vortex of much higher strength; and more generally, $N+K$ configurations. We also give examples of configurations that have not been studied experimentally, including $N+2$ configurations, where $N$ vortices aggregate inside an ellipse. Finally, we introduce an artificial 'damping' to the vortex dynamics, in an attempt to explain the phenomenon of crystallization that is often observed in real experiments. The diffusion breaks the conservative structure of vortex dynamics, so that any initial conditions converge to the lattice-like relative equilibrium.## 1. Introduction

The dynamics of $N$ point vortices in a plane is a classical problem that goes back to the works of Helmholtz [1], who first described this model as a fluid analogue to the $N$-body problem in celestial mechanics. Fundamentally, point vortices correspond to singularities in an ideal irrotational flow. These singularities in turn characterize the flow itself. Vortex dynamics have been reproduced in many physical experiments, starting with those of Meyer [2] of floating needle magnets in an applied magnetic

Royal Society Publishing
Informing the science of the future

© 2013 The Author(s) Published by the Royal Society. All rights reserved.

field. Yarmchk *et al.* [3] obtained stable vortex lattices in a rotating $^4$He superfluid. Vortex dynamics have also been observed in Bose–Einstein condensates [4–6] and in electron columns confined in a Malmberg–Penning trap [7], which allow for a high degree of control over initial vortex configurations [8]. Vortex dynamics were also reproduced experimentally using a system of small rotating discs [9–11]. See reference [12] for a recent review of these experiments. Some examples of vortices in nature include geophysical flows [13] such as Jupiter's red spot [14], intensifying hurricanes [15] and plasma flows [16].

As first described by Helmholtz, each vortex generates a rotational velocity field that advects all other vortices. This yields a system of ordinary differential equations (ODEs) for vortex centres $z_j$ which we will refer to as the *vortex model*,

$$\frac{dz_j}{dt} = i \sum_{k \neq j} \gamma_k \frac{z_j - z_k}{|z_j - z_k|^2}, \quad j = 1 \ldots N. \tag{1.1}$$

Here, $i = \sqrt{-1}$, $z_k \in \mathbb{C}$ and $2\pi\gamma_k$ is the circulation of the $k$th vortex; we will assume throughout the paper that $\gamma_k > 0$. There is an extensive literature that describes vortex dynamics of a small number of vortices, or special vortex configurations, such as vortices arranged on a ring (see [17–19] for the overview of the field). The vortex model is a Hamiltonian system of many variables. As such, general initial conditions with four or more vortices typically result in chaotic dynamics. However, there are many special arrangements, which lead to the rigidly rotating configurations of vortices, called *relative equilibria* [12,17,20,21]. These equilibria can be either stable [12,22] or unstable [23]. For example, $N$ vortices of equal strength arranged uniformly along a ring is a basic relative equilibrium, which is stable for $N \leq 6$ and unstable for $N \geq 8$, with $N = 7$ being the threshold case [24–26]. The stable relative equilibria are of particular physical importance, and are often observed in experiments, even when starting with arbitrary initial conditions. In real experiments, there is usually some form of dissipation present, which disturbs the underlying Hamiltonian structure of vortex dynamics and, in practice, causes a decay towards a stable lattice-type relative equilibrium over a long time. For example, this process was observed in experiments with magnetized electron columns [7,8,27] and has been dubbed 'crystallization' in [8].

Motivated by rotating $^4$He superfluid experiments [3], Campbell & Ziff [22] classified many stable configurations for a *small* number of vortices of equal strength. Our goal here is to describe the stable configuration in the continuum limit of a *large* number of vortices $N \to \infty$. The key observation is that the model (1.1) is intimately connected to the following *aggregation model*,

$$\frac{dx_j}{dt} = \sum_{k \neq j} \gamma_k \frac{x_j - x_k}{|x_j - x_k|^2} - \omega x_j. \tag{1.2}$$

Here, $\omega$ is the angular velocity of the relative equilibrium. Model (1.2) was recently investigated in studies [28,29] in connection with swarm formations in biology. There is a one-to-one correspondence between the steady states $x_j(t) = \xi_j$ of (1.2) and the relative equilibrium $z_j(t) = e^{\omega i t} \xi_j$ of (1.1). Moreover, as we show below, this correspondence carries over to stability: the equilibrium $x_j(t) = \xi_j$ is asymptotically stable for the aggregation model (1.2) if and only if the relative equilibrium $z_j(t) = e^{\omega i t} \xi_j$ is stable (neutrally, in the Hamiltonian sense) for the vortex model (1.1). While the steady states and their local stability are the same for the vortex and aggregation model, the latter has much simpler dynamics, and many results can be explicitly obtained in the large-$N$ limit as we show below.

Throughout the paper, we make use of the fact that $\omega > 0$, which follows from our original assumption that $\gamma_j > 0$ as we now show. Indeed, we have

$$\omega = \frac{\sum \sum_{j \neq k} \gamma_k \gamma_j}{\sum_j \gamma_j |\xi_j|^2}, \tag{1.3}$$

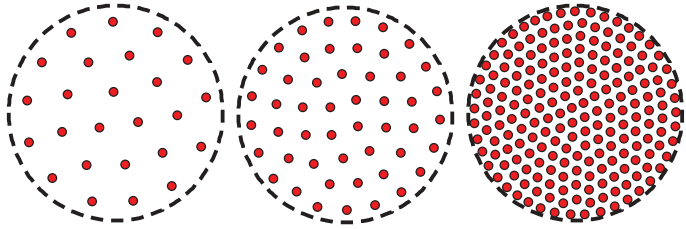

**Figure 1.** Stable relative equilibria of $N = 25, 50$ and $200$ vortices of equal strength. The dashed line shows the analytical prediction $R_0 = \sqrt{N\gamma/\omega}$ of the swarm radius in the $N \to \infty$ limit (see (2.3)). (Online version in colour.)

where $x_j(t) = \xi_j$ is the equilibrium state of (1.2) corresponding to the relative equilibrium $z_j(t) = \exp(i\omega t)\xi_j$ of (1.1). This can be seen by multiplying the right-hand side of (1.2) by $\gamma_j \bar{\xi}_j$, summing over $j$, and setting the sum to zero. The quantity $\sum\sum_{j\neq k}\gamma_k\gamma_j$ is the total angular vortex momentum, and $\sum_j \gamma_j |\xi_j|^2$ is the angular impulse.[1]

## 2. Equilibrium and stability of vortices of equal strength

Consider the special case of $N$ vortices of equal strength $\gamma_k = \gamma$ so that the aggregation model (1.2) becomes

$$\frac{dx_j}{dt} = \frac{1}{N} \sum_{\substack{k=1\ldots N \\ k\neq j}} a\frac{x_j - x_k}{|x_j - x_k|^2} - x_j, \quad j = 1\ldots N, \qquad (2.1)$$

where $a = N\gamma_k/\omega$ and where we rescaled the time $t \to \omega^{-1}t$. The distinguished $1/N$ scaling makes it possible to take the mean-field limit $N \to \infty$ which yields a non-local partial differential equation (PDE) [28–30]

$$\rho_t + \nabla \cdot (\rho v) = 0; \quad v(x) = a\int_{\mathbb{R}^2} \rho(y) \frac{x-y}{|x-y|^2} dy - x. \qquad (2.2)$$

Here, $\rho(x,t)$ approximates the particle density normalized so that $\int_D \rho(x,t)\,dx$ represents the fraction of particles inside a domain $D$ with $\int_{\mathbb{R}^2} \rho(x,t) = 1$.

The system (2.2) was analysed in detail in recent studies [28,29]. It was shown that in the limit $t \to \infty$, the density $\rho(x,t)$ approaches a steady state which is constant inside a disc of radius $\sqrt{a}$ and is zero outside such a disc: $\rho(x,t) \to 1/(a\pi)$ if $|x| < \sqrt{a}$ and $\rho(x,t) \to 0$ as $t \to \infty$ otherwise. For the vortex model, this result implies that for large $N$, the stable relative equilibrium for (1.1) with $\gamma_k = \gamma$ consists of particles uniformly distributed inside a disc of radius

$$R_0 = \sqrt{\frac{N\gamma}{\omega}}. \qquad (2.3)$$

The radius $R_0$ was previously derived in the physics literature, see for example [31] and in [32, eqn. 25]. The stability can also be ascertained: as was shown in [29], the uniform disc is the global attractor for the continuum limit (2.2) of the aggregation model. An immediate consequence is that the uniform disc is the only stable relative equilibrium of the vortex model (1.1) with $\gamma_k = \gamma$ in the large $N$ limit. While in the derivation of (2.3), we assumed that $R_0$ is $O(1)$, the formula actually holds even when $\gamma/\omega = O(1)$. This is because one can always rescale the space to make $R_0$ of $O(1)$. Indeed, the rescaling $x_j = \sqrt{N/a}\tilde{x}_j$, $t = \sqrt{N/a}\tilde{t}$ eliminates $a/N$ from equation (2.3); hence, these results are independent of scaling.

Figure 1 shows that the continuum limit radius (2.3) provides a very good estimate of the equilibrium radius even for relatively small $N = 25$; as expected, larger $N$ results in an even better agreement. Below, we will re-derive this result as a special case of the $N + 1$ configuration.

---
[1]We thank the anonymous referee for pointing out formula (1.3) to us.

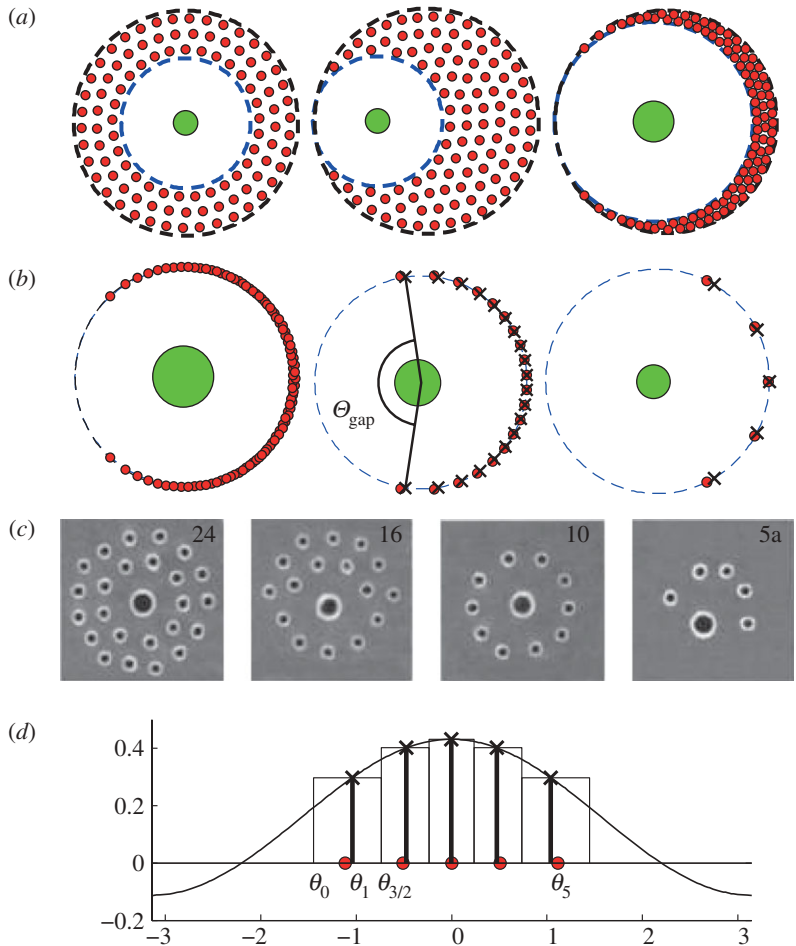

**Figure 2.** Stable relative equilibria for $N+1$ system of vortices (3.1) and (3.2). (*a*): $N = 100$, $b = 1$ are fixed with $a = 2, 2$ and 0.25 (left to right). The asymptotic predictions are indicated by dashed curves (result 1). (*b*): parameters $a = 0.05$, $b = 1$ are fixed with $N = 100, 20$ and $5$ (left to right). Asymptotic prediction (result 3) is indicated by crosses middle-right figure. (*c*): experiments with small, floating magnetized discs from [9]. Reprinted with permission from the authors. Copyright (2001) by the American Physical Society. (*d*): the angular density distribution of particles is well-approximated by (3.8) shown by crosses; circles are the full numerics. $\alpha$ is as given in result 3. (Online version in colour.)

## 3. $N+1$ problem

A well-studied case in the literature going back to Havelock [25] consists of a ring of $N$ vortices of equal strength surrounding a single vortex of a much higher strength. This is the fluids analogue of the 'rings of saturn' problem in celestial mechanics that was first studied by Maxwell [33]. Cabral & Schmidt [26] established the stability of an $N+1$ uniform ring, provided that the strength of the central vortex is between a certain lower and an upper bound. Below the lower bound, the ring has high-mode instabilities that cause the ring to deform into an annulus. This is illustrated in the left-top panel of figure 2. Above the upper bound, there is a low-mode instability that causes the ring to deform into a stable non-uniform ring-like state [20], as illustrated in the second row of figure 2. The non-uniform state of this type was also observed experimentally in [9] and in [8] (figure 2). For the experiments with Bose–Einstein condensates, it has been shown that higher charge vortices are both experimentally and theoretically potentially unstable, owing to break-up to lower topological charges [34–36], which makes the $N+1$ configurations more difficult to achieve.

In this section, we will construct the $N+1$ states in the large $N$ limit. Label $\eta = x_{N+1}$ and rewrite (1.2) as

$$\frac{dx_j}{dt} = \frac{a}{N} \sum_{\substack{k=1...N \\ k \neq j}} \frac{x_j - x_k}{|x_j - x_k|^2} + b \frac{x_j - \eta}{|x_j - \eta|^2} - x_j, \quad j = 1...N, \quad (3.1)$$

and

$$\frac{d\eta}{dt} = \frac{a}{N} \sum_{k=1...N} \frac{\eta - x_k}{|\eta - x_k|^2} - \eta_j, \quad (3.2)$$

where $a = N\gamma_k/\omega$, $k = 1...N$; $b = \gamma_{N+1}/\omega$ and where we rescaled the time $t \to \omega^{-1}t$. Taking the mean-field limit $N \to \infty$ as before yields a PDE for the density distribution of small vortices $\rho(y)$ which satisfies

$$\left.\begin{aligned} \rho_t + \nabla \cdot (\rho v) &= 0; \\ v(x) &= a \int_{\mathbb{R}^2} \rho(y) \frac{x-y}{|x-y|^2} dy + b \frac{x-\eta}{|x-\eta|^2} - x \\ \frac{d\eta}{dt} &= a \int_{\mathbb{R}^2} \rho(y) \frac{\eta - y}{|\eta - y|^2} dy - \eta. \end{aligned}\right\} \quad (3.3)$$

We make the following ansatz for the steady state:

$$\rho(x) = \begin{cases} \rho_0, & \text{if } x \in B_{R_0}(0) \setminus B_{R_1}(\eta) \\ 0, & \text{otherwise}, \end{cases}$$

where $B_R(z)$ denotes a disc of radius $R$ centred at $z$; $R_0, R_1 \in \mathbb{R}^+$ and $\eta \in B_{R_0}(0)$ are chosen such that $B_{R_1}(\eta) \subset B_{R_0}(0)$. The normalization $\int_{\mathbb{R}^2} \rho(x) dx = 1$ yields $\rho_0 \pi (R_0^2 - R_1^2) = 1$. Using the identity

$$\int_{B_R(z)} \frac{x-y}{|x-y|^2} dy = \begin{cases} \pi(x-z), & x \in B_r(z) \\ \pi R^2 \frac{x-z}{|x-z|^2}, & x \notin B_r(z), \end{cases}$$

we find that for $x \in B_{R_0}(0) \setminus B_{R_1}(\eta)$,

$$v(x) = (a\rho_0\pi - 1)x + (-a\rho_0\pi R_1^2 + b)\frac{x-\eta}{|x-\eta|^2}; \quad \eta_t = (a\rho_0\pi - 1)\eta.$$

At the steady state of (3.3), $v(x) = 0 = \eta_t$ so that $a\rho_0\pi = 1$; $a\rho_0\pi R_1^2 = b$. Solving for $R_1, R_0$, we obtain the following result.

**Result 3.1.** Define $R_1 = \sqrt{b}$, $R_0 = \sqrt{a+b}$ and suppose that $\eta$ is any point such that $B_{R_1}(\eta) \subset B_{R_0}(0)$. Then, the equilibrium solution for (3.3) is constant inside $B_{R_0}(0) \setminus B_{R_1}(\eta)$ and is zero outside.

This result is illustrated in figure 2, first row, where the boundaries of discs $B_{R_1}(\eta), B_{R_0}(0)$ are shown by dashed lines. Excellent agreement between the continuum limit result is observed. Unlike the $N+0$ problem, the relative equilibrium for the $N+1$ problem is non-unique: any choice of $\eta$ yields a steady state as long as $|\eta| < R_0 - R_1$.

In the limit $a \to 0$, note that $R_1 \to R_0$ so that the steady-state approaches a (possibly non-uniform) ring solution. Assuming that $\eta \sim 0$ is at its equilibrium, it was shown in [20] that the evolution of the small vortices is given by $x_j(t) \sim R_0 e^{i\theta_j(t)}$ where $\theta_j$ satisfies, after a rescaling $t \to a^{-1}t$,

$$\frac{d\theta_j}{dt} = \frac{1}{N} \sum_{k \neq j} \left( \frac{\sin(\theta_j - \theta_k)}{2 - 2\cos(\theta_j - \theta_k)} - \sin(\theta_j - \theta_k) \right). \quad (3.4)$$

In the mean-field limit $N \to \infty$, the density distribution $\rho(\theta)$ for the angles $\theta_j$ satisfies

$$\left.\begin{aligned} \rho_t + (\rho v_\theta)_\theta &= 0, \\ v(\theta) &= \text{PV} \int_{-\pi}^{\pi} \rho(\phi) \left( \frac{\sin(\theta - \phi)}{2 - 2\cos(\theta - \phi)} - \sin(\theta - \phi) \right) d\phi, \end{aligned}\right\} \quad (3.5)$$

and

where PV denotes the principal value integral, and $\int_{-\pi}^{\pi} \rho = 1$. Using a formal expansion

$$\frac{\sin t}{2 - 2\cos t} - \sin t = \sin(2t) + \sin(3t) + \sin(4t) + \ldots, \quad (3.6)$$

note that up to rotations, the steady-state density $\rho(\theta)$ for which $v = 0$ must be of the form

$$\rho(\theta) = \frac{1}{2\pi}(1 + \alpha \cos \theta). \quad (3.7)$$

This formula was first derived in [37] using a similar technique. The free constant $\alpha$ corresponds to the non-uniqueness of the steady state of the continuum limit (3.5) of the $N+1$ problem. This steady state is in fact stable (see appendix A), and the density is strictly positive whenever $|\alpha| < 1$. However, for any *finite* $N$, the density $\rho$ is quantized; this has the effect of choosing a specific constant $\alpha$. The discrete locations are well approximated by

$$\int_{-\pi}^{\theta_j} \frac{1}{2\pi}(1 + \alpha \cos \theta)\, d\theta = \frac{(j - 1/2)}{N} \quad (3.8)$$

(figure 2, row 4). The relative equilibrium exhibits a gap along the unit circle. In appendix A, we show the following result.

**Result 3.2.** In the limit $a \ll 1$ and $N \gg 1$, the equilibrium $\theta_j$ for the problem (3.4) is approximated by solving (3.8), where $\alpha \sim 1 + A\ N^{-1/3}$ with $\theta_1 \sim -\pi + B\ N^{-1/3}$, where the constants $A \approx 2.0699802$ and $B \approx 4.122044$ satisfy equations (A 4). The gap in the steady state, given by $\Theta_{\text{gap}} = 2(\theta_1 + \pi)$, is of the size

$$\Theta_{\text{gap}} = 2BN^{-1/3} \approx 8.244N^{-1/3}. \quad (3.9)$$

This result is illustrated in figure 2, second row. Despite the low-power scaling $O(N^{-1/3})$ for the gap, formula (3.9) is very effective even for relatively few vortices $N = 5$.

## 4. $N + K$ problem

It is straightforward to generalize result 1 to the situation where there are $K$ large vortices and $N$ small ones. In analogy to the $N+1$ case, we let $a = N\gamma_k/\omega, k = 1 \ldots N$; $b_k = \gamma_{N+k}/\omega, k = 1 \ldots K$ and rescale the time $t \to \omega^{-1}t$. The mean-field limit $N \to \infty$ for the velocity $v$ of the density distribution $\rho$ of vortices then becomes a coupled system

and
$$\left.\begin{aligned}v(x) &= a\int_{\mathbb{R}^2} \rho(y)\frac{x-y}{|x-y|^2}dy + \sum_{k=1\ldots K} b_k \frac{x - \eta_k}{|x - \eta_k|^2} - x, \\ \frac{d\eta_j}{dt} &= a\int_{\mathbb{R}^2} \rho(y)\frac{\eta_k - y}{|\eta_k - y|^2}dy + \sum_{\substack{k=1\ldots K \\ k \neq j}} b_k \frac{\eta_j - \eta_k}{|\eta_j - \eta_k|^2} - \eta_j, \quad j = 1 \ldots K.\end{aligned}\right\} \quad (4.1)$$

The generalization of result 1 then yields

**Result 4.1.** Let $R_k = \sqrt{b_k}, k = 1 \ldots K$ and $R_0 = \sqrt{a + b_1 + \cdots + b_K}$. Suppose $\eta_1 \ldots \eta_K$ are such $B_{R_1}(\eta_1) \ldots B_{R_K}(\eta_K)$ are all disjoint and are contained inside $B_{R_0}(0)$. The equilibrium density for (4.1) is constant inside $B_{R_0}(0) \setminus \bigcup_{k=1}^{K} B_{R_k}(\eta_k)$ and is zero outside.

In analogy to the $N + 1$ problem, such an equilibrium has $K$ free parameters $\eta_1 \ldots \eta_K$. Some examples are given in figure 3. Unlike the $N+1$ problem, not all values of parameters are allowed in result 3. For example, for the $N + 2$ problem, such solutions exist if and only if $R_1 + R_2 < R_0$ or $\sqrt{b_1 b_2} \leq a/2$. In the opposite case, the two smaller discs no longer fit inside the big disc; this is illustrated in figure 3, row 2.

When $K > 2$ and in the limit $a \to 0$, the integral terms in (4.1) disappear at the leading order, so that $K$ large vortices 'decouple' from the $N$ small vortices, and their behaviour is simply given by a lower-dimensional $K$-vortex problem obtained by setting $a = 0$. The $N$ small equilibria aggregate at several points as illustrated in figure 3 (middle left and bottom). These special points correspond to the stagnation points of the velocity field (4.1) with $a = 0$. As we now show, the shape of the swarm around these points can be computed explicitly. For simplicity, we concentrate on a special case of two big vortices of equal strength $b_1 = b_2 = b$ and one aggregation of small vortices. This configuration is shown in figure 3, middle right panel. The relative equilibrium position of the two big vortices $\eta_1, \eta_2$ satisfies

$$b\frac{\eta_2 - \eta_1}{|\eta_2 - \eta_1|^2} = \eta_1 = -\eta_2.$$

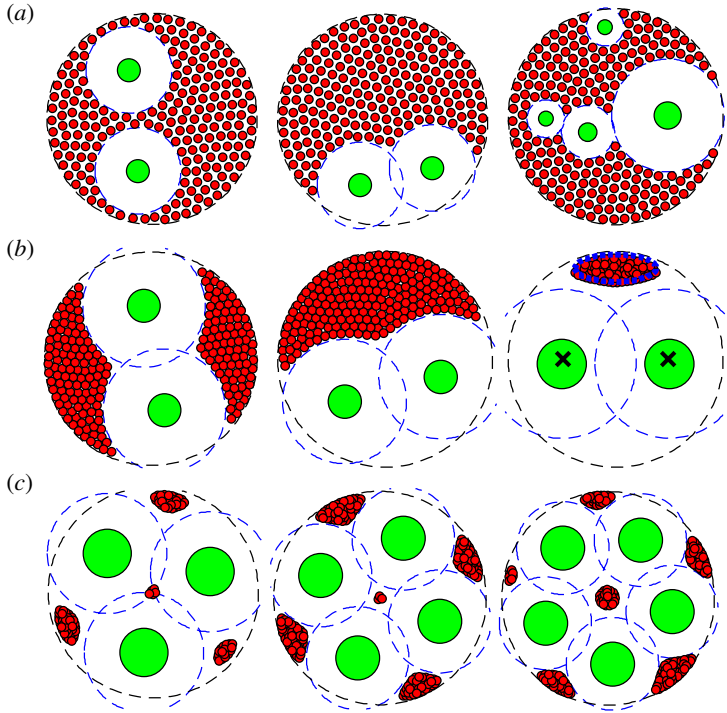

**Figure 3.** $N + K$ relative equilibria. Small dashed circles are the large-$N$ limit result $B_{R_1}(\eta_1) \ldots B_{R_K}(\eta_K)$ and big dashed circle has the radius $R_0$, see result 3. For all snapshots, $N = 200$ and $K$ corresponds to the number of big dots. (a) Left and middle: $a = 1$, $(b_1, b_2) = (0.25, 0.5)$; right: $a = 1$, $(b_1 \ldots b_4) = (0.05, 0.05, 0.1, 0.45)$. (b) $b_1 = b_2 = 1$ with $a = 1$ (left and centre) and $a = 0.1$ (right). (c) $b_j = 1$ and $a = 0.1, 0.3, 0.3$ resp. for left, centre and right. (b) Right: the thick dashed curve is the predicted ellipse shape. Crosses are the asymptotic result $\eta_1 = 1/\sqrt{2} = -\eta_2$. (Online version in colour.)

By rotation, assume $\eta_1, \eta_2$ are on the $x$-axis so that $\eta_1 = \sqrt{b/2} = -\eta_2$. For the small vortices and in the limit $a \to 0$, we set $x_j = X + a^{1/2}\xi_j + O(a)$. Collecting the $O(1)$ terms, we then obtain

$$b\left(\frac{X - \eta_1}{|X - \eta_1|^2} + \frac{X - \eta_2}{|X - \eta_2|^2}\right) - X = 0.$$

This yields $X = \pm\sqrt{\frac{3}{2}b}\mathrm{i}$. At the next order, collecting the $O(a^{1/2})$ terms, we then obtain system

$$\frac{d\xi_j}{dt} = \frac{1}{N}\sum_{\substack{k=1\ldots N \\ k \neq j}} \frac{1}{\bar{\xi}_j - \bar{\xi}_k} + \frac{1}{2}\bar{\xi}_k - \xi_k \tag{4.2}$$

where the bar denotes the complex conjugate. The system (4.2) was recently analysed in [38] in the context of singularly perturbed aggregation kernels. Using a complex variables method, it was shown that $\xi_j$ concentrate uniformly inside an *ellipse* whose radii are $\sqrt{3}$ and $\sqrt{1/3}$ and whose major axis is parallel to the $x$-axis. This ellipse, along with the full solution, is plotted in figure 3, right-middle panel; good agreement is observed.

## 5. Crystallization

Owing to the conservative nature of the idealized vortex dynamics (1.1), random initial conditions typically result in chaotic motion, and stable relative equilibria are only neutrally stable: small perturbations neither grow nor decay. By contrast, in experiments [3,7–9], arbitrary initial conditions often converge to an asymptotically stable lattice-like state. Similar stable lattices have also been observed in Bose–Einstein condensates [4]. The emergence of such lattices can be explained by adding a small 'damping' term that destroys the Hamiltonian structure. Here, we propose the following phenomenological model that incorporates this friction:

$$\frac{dz_j}{dt} = i\sum_{k\neq j} \gamma_k \frac{z_j - z_k}{|z_j - z_k|^2} + \mu\left(\sum_{k\neq j} \gamma_k \frac{z_j - z_k}{|z_j - z_k|^2} - \omega z_j\right). \quad (5.1)$$

The constant $\mu \geq 0$ models damping effects. The term $\gamma_k(i+\mu)(z-z_k/|z-z_k|^2)$ corresponds to a velocity field generated by an outwards spiralling source that adds local repulsion of strength $\mu$. The term $-\mu\omega z_j$ keeps the vortices confined near the origin; it arises naturally for vortices confined to a circular domain via boundary effects. The original vortex model (1.1) is a special case of (5.1) obtained by setting $\mu = 0$. Moreover, the aggregation model (1.2) is also a special case of (5.1) by taking $\mu \to \infty$ after rescaling the time $t \to \mu^{-1}t$.

The specific form of (5.1) is motivated by the fact that any relative equilibrium $z_j(t) = \exp(\omega it)\xi_j$ of (5.1) is also a relative equilibrium of (1.1) for any $\mu$; and vice versa. Furthermore, we now show that the relative equilibrium $z_j(t) = \exp(\omega it)\xi_j$ of the vortex model (1.1) is stable if and only if it is a stable for the damped system (5.1), and if and only if the corresponding equilibrium $x_j(t) = \xi_j$ of the aggregation model (1.2) is stable. Note that systems (1.1), (1.2) and (5.1) are all invariant under rotations; hence, there is always a zero eigenvalue of the relative equilibrium that corresponds to the rotation invariance. So, we define stability to mean that all *other* eigenvalues are non-positive. Because the vortex model (1.1) is Hamiltonian, its eigenvalues come in pairs $\pm\lambda$. The (neutral) stability in this case means that all eigenvalues (except for the zero rotational mode) are purely imaginary.

We linearize around the relative equilibrium by setting $z_j(t) = \exp(\omega it)(\xi_j + \eta_j(t))$ where $\eta \ll 1$. We then obtain the $2N \times 2N$ linear system of ODEs

$$\frac{d}{dt}\eta = (i+\mu)(L\bar{\eta} - \omega\eta), \quad \frac{d}{dt}\bar{\eta} = (-i+\mu)(\bar{L}\eta - \omega\bar{\eta}), \quad (5.2)$$

where the overbar denotes complex conjugation; $\eta = (\eta_1, \ldots \eta_N)$ and $L$ is an $N \times N$ matrix with $L_{jk} = \gamma_k/(\bar{\xi}_j - \bar{\xi}_k)^2$ for $j \neq k$ and with $L_{jj} = -\sum_{k\neq j} L_{jk}$.

Eliminating $\bar{\eta}$ from (5.2), we obtain

$$\eta_{tt} = (i+\mu)L\bar{\eta}_t - (i+\mu)\omega\eta_t$$
$$= (1+\mu^2)L\bar{L}\eta - (-i+\mu)\omega(i+\mu)L\bar{\eta} - (i+\mu)\omega\eta_t$$
$$= (1+\mu^2)(L\bar{L} - \omega^2)\eta - 2\mu\omega\eta_t. \quad (5.3)$$

Setting $\eta(t) = \exp(\lambda t)v$ in (5.3), we find that $\lambda$ satisfies

$$\lambda^2 + 2\mu\omega\lambda + (\omega^2 - \sigma)(\mu^2 + 1) = 0, \quad (5.4)$$

where $\sigma$ is an eigenvalue of $L\bar{L}$ with corresponding eigenvector $v$. Note that $L = D\hat{L}$, where $D = \text{diag}(\gamma_1^{-1} \ldots \gamma_N^{-1})$ is a positive diagonal matrix, and $\hat{L}$ is a symmetric matrix. It follows that $L\bar{L} = D^{1/2}(D^{1/2}\hat{L}D^{1/2}\overline{D^{1/2}\hat{L}D^{1/2}})D^{-1/2}$ is similar to the hermitian matrix $D^{1/2}\hat{L}D^{1/2}\overline{D^{1/2}\hat{L}D^{1/2}}$, so that $\sigma \in \mathbb{R}^+$. Hence, recalling the assumption $\mu, \omega > 0$, we obtain that $Re(\lambda) < 0$ if and only if $\sigma < \omega^2$.

The original vortex model (1.1) corresponds to taking $\mu = 0$ so that $\lambda = \pm\sqrt{\sigma - \omega^2}$. If $\sigma < \omega^2$, then $\lambda$ is purely imaginary which corresponds to the neutral stability. Otherwise, $\lambda = +\sqrt{\sigma - \omega^2}$ is a positive eigenvalue, and the relative equilibrium is unstable.

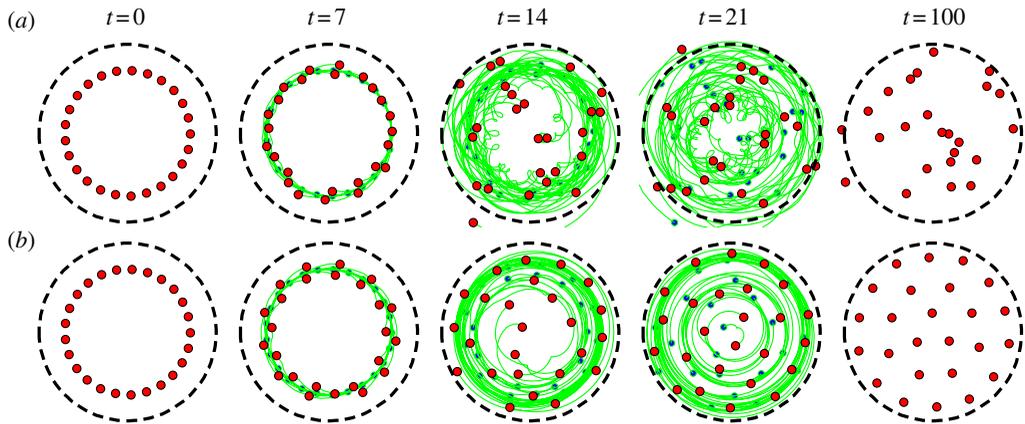

**Figure 4.** The effect the damping $\mu$ in (5.1) on the vortex dynamics. Snapshots of the dynamics are shown with time $t$ as indicated. (a) $\mu = 0$. (b) $\mu = 0.1$. Initial conditions and all other parameters are the same for the two runs. We invite the reader to see the full movies at this address: http://www.mathstat.dal.ca/~tkolokol/vortexswarms. (Online version in colour.)

An analogous computation for the aggregation model (1.2) shows that the eigenvalues of the steady state $x_j(t) = \xi_j$ are similarly given by

$$\lambda^2 + 2\omega\lambda + (\omega^2 - \sigma) = 0 \qquad (5.5)$$

so that $\lambda = -\omega \pm \sqrt{\sigma} \in \mathbb{R}$ with both $\lambda < 0$ if and only if $\sigma < \omega^2$ (note that (5.5) can also obtained from (5.4) by taking the limit $\mu \to \infty$ after rescaling $\lambda \to \mu\lambda$). This establishes the equivalence of linear stability of the steady state $x_j(t) = \xi_j$ of the aggregation model (1.2), the neutral stability of the corresponding relative equilibrium $z_j(t) = e^{i\omega t}\xi_j$ of the vortex model (1.1), and the linear stability of the model with damping (5.1). We remark that this method is similar to the argument that was used in [39].

## 6. Discussion

Vortex dynamics is a very old subject, dating back to the 1850s. In this paper, we explored the connection between vortex dynamics and a model of biological swarming. This connection allows us to obtain many new as well as existing results in the large $N$ limit. For example, previous works on the $N + 1$ configurations [20,26] consider only the case where the $N$ vortices form a ring around the big vortex. Here, we treat the more general case where the equilibrium is bounded between two circles. The previously known cases can be recovered as limiting case when the radius of the inner circle approaches the radius of the outer circle. Moreover, we also computed asymptotically the gap in the $N + 1$ non-uniform ring using a similar approach. This result is also new.

Many open questions remain. The stability of $N + K$ configurations has not been studied, except when $K = 0$ or for a very restricted case of the $N + 1$ problem (see appendix A). For the $N + 2$ problem, the steady state as described in result 2 requires that the two discs inside the swarm are disjoint. But there are other configurations, such as those illustrated in figure 3, top-middle panel, where the two discs overlap by an $O(1)$ amount. Numerical simulations indicate that the amount of the overlap is independent of initial conditions, provided that the initial conditions are such that the small vortices are all on one side of the two big vortices.

Numerical simulations of the crystallization model (5.1) indicate that the steady state attained in figure 4 is a global attractor. This remains to be shown. Model (5.1) is the simplest possible model that incorporates damping while still preserving the relative equilibrium of the original vortex model (1.1). It would be interesting to incorporate the damping effects in a more systematic way starting from first principles. See [40] for work in this direction.

We thank the anonymous referees for a careful reading, very useful suggestions including formula (1.3), and for finding several mistakes in the first draft of the paper. Their comments helped to improve the paper significantly. We thank James von Brecht for fruitful discussions. T.K. was supported by a grant from AARMS CRG in Dynamical Systems and NSERC grant no. 47050. Some of the research for this project was carried out while Y.C. and T.K. were supported by the California Research Training Programme in Computational and Applied Mathematics (NSF grant no. DMS-1045536).

## Appendix A

### (a) Stability of the $N+1$ solution (3.7)

We now show that (3.7) is a stable steady state of the continuum equation (3.5). For simplicity, we restrict the discussion to symmetric solutions that have the Fourier expansion

$$\rho(t) = C_0 + \sum_{m=1}^{\infty} 2C_m(t)\cos(m\theta).$$

Using (3.6), the velocity is then given by

$$v = \sum_{m=2}^{\infty} 2\pi C_m \sin(m\theta).$$

Substituting into $\rho_t + (\rho v_\theta)_\theta = 0$, we obtain the following infinite system of ODEs for the Fourier coefficients

$$\left.\begin{aligned}\frac{d}{dt}C_1 &= -\pi C_1 C_2, \\ \frac{d}{dt}C_2 &= -2\pi(C_0 C_2 + C_1 C_3) \\ \text{and} \quad \frac{d}{dt}C_m &= -m\pi(C_0 C_m + C_1(C_{m-1} + C_{m+1})), \quad m \geq 3.\end{aligned}\right\} \quad (A\,1)$$

This system admits a steady state $C_m = 0$, $m \geq 2$ with $C_1$ arbitrary. Linearizing about this state yields the infinite Jacobian matrix

$$\frac{J}{\pi} = \begin{bmatrix} 0 & -C_1 & & & \\ 0 & -2C_0 & -2C_1 & & \\ 0 & -3C_1 & -3C_0 & -3C_1 & \\ & & -4C_1 & -4C_0 & -4C_1 \\ & & & \ddots & \ddots & \ddots \end{bmatrix}.$$

Applying the Gershgorin Theorem to the columns of $J$, its eigenvalues are non-positive provided that $|C_1| < C_0/2$, which is precisely the condition that steady-state density $\rho(\theta) = C_0 + 2C_1 \cos\theta$ is positive.

### (b) Derivation of result 2

The continuum density $\rho(\theta) = (1/2\pi)(1 + \alpha\cos\theta)$ is approximated by a discrete density $\rho_N = (1/N)\sum_{k=1}^{N} \delta(\theta - \theta_k)$, where $\theta_k$ are given by (3.8). The idea is to choose $\alpha$ such that the velocity at the boundary is zero: $d\theta_1/dt = 0$, where $d\theta_1/dt$ is given by (3.4). We then estimate the discrete sum in (3.4) by $(d\theta_1/dt) \sim$

$$\int_{\theta_{1+1/2}}^{\pi} \frac{1}{2\pi}(1 + \alpha\cos\phi)\left(\frac{\sin(\theta_1 - \phi)}{2 - 2\cos(\theta_1 - \phi)} - \sin(\theta_1 - \phi)\right)d\phi,$$

where $\theta_{1+1/2}$ is given by (3.8) with $j = 1 + 1/2$; refer to the bottom of figure 2. This integral can be evaluated exactly using the identities

$$\int \sin t \left(\frac{\sin(t)}{2 - 2\cos(t)} - \sin(t)\right) dt = \frac{1}{2}(\sin t)(\cos t + 1);$$

$$\int \cos t \left( \frac{\sin(t)}{2 - 2\cos(t)} - \sin(t) \right) dt = \ln(1 - \cos(t)) + \frac{\cos t}{2}(\cos t + 1).$$

We then expand using the following scaling:

$$\alpha = 1 + AN^{-1/2}; \quad x_1 = -\pi + BN^{-1/3}; \quad x_{1+1/2} = -\pi + CN^{-1/3}$$

where $A, B, C$ are $O(1)$. Setting $d\theta_1/dt = 0$ and expanding in Taylor series yields, at the leading order,

$$C(2B + C) = 2\ln\left(\frac{C}{B-1}\right)(2A - B^2). \tag{A 2}$$

The expressions (3.8) for $x_1$ and $x_{1+1/2}$ yield to leading order,

$$\frac{B^3}{12} - \frac{AB}{2} = \frac{\pi}{2}; \quad \frac{C^3}{12} - \frac{AC}{2} = \pi. \tag{A 3}$$

Together, (A 2) and (A 3) give a system of three algebraic equations for $A, B, C$ independent of $N$ to leading order. Two of the variables can be eliminated by defining $u = C/B$; $v = A/B^2$. Eliminating $v$, we then obtain

and
$$\left.\begin{array}{c} (8 - 6u + 2u^3)\ln(u - 1) = 3u(u^2 - 4) \\ B^3 = \dfrac{6\pi(2 - u)}{u(u^2 - 1)}; \quad A = B^2 \dfrac{1}{6}\dfrac{2 - u^3}{2 - u}. \end{array}\right\} \tag{A 4}$$